\documentclass[traditabstract]{aa} 
\usepackage{graphicx}
\usepackage{txfonts}
\usepackage{ifthen}
\usepackage{natbib}
\bibpunct{(}{)}{;}{a}{}{,}

\newcommand\degree{\degr}
\newcommand\degrees\degree
\newcommand\vs{{\em vs.}}

\DeclareSymbolFont{UPM}{U}{eur}{m}{n}
\DeclareMathSymbol{\umu}{0}{UPM}{"16}
\let\oldumu=\umu
\renewcommand\umu{\ifmmode\oldumu\else\math{\oldumu}\fi}
\newcommand\micro{\umu}
\newcommand\microns{\micro m}

\let\oldsim=\sim
\renewcommand\sim{\ifmmode\oldsim\else\math{\oldsim}\fi}
\let\oldpm=\pm
\renewcommand\pm{\ifmmode\oldpm\else\math{\oldpm}\fi}
\newcommand\by{\ifmmode\times\else\math{\times}\fi}

\newcommand\psfr{\ifmmode{F\sb{p}/F\sb{s}}\else \math{F\sb{p}/F\sb{s}}\fi }

\newcommand\twoseventeenb{HD~217107\,b}

\newcommand\ttt[1]{\ifmmode 10\sp{#1}   \else \math{10\sp{#1}}\fi}
\newcommand\tttt[1]{\by\ttt{#1}}

\newbox{\wdbox}
\renewcommand\c{\setbox\wdbox=\hbox{,}\hspace{\wd\wdbox}}
\renewcommand\i{\setbox\wdbox=\hbox{i}\hspace{\wd\wdbox}}

\newcount\timect
\newcount\hourct
\newcount\minct
\newcommand\now{\timect=\time \divide\timect by 60
         \hourct=\timect \multiply\hourct by 60
         \minct=\time \advance\minct by -\hourct
         \number\timect:\ifnum \minct < 10 0\fi\number\minct}

\newcommand\mctc{\multicolumn{2}{c}}

\catcode`@=11

\newcommand\comment[1]{}

\renewcommand\math[1]{$#1$}

\comment{alignment tab}
\let\atab=&

\let\oldmsp=\sp
\let\oldmsb=\sb
\renewcommand\sp[1]{\ifmmode
	   \oldmsp{#1}%
	 \else\strut\raise.85ex\hbox{\scriptsize #1}\fi}
\renewcommand\sb[1]{\ifmmode
	   \oldmsb{#1}%
	 \else\strut\raise-.54ex\hbox{\scriptsize #1}\fi}
\newcommand\msp[1]{\ifmmode
	   \oldmsp{#1}
	 \else \math{\oldmsp{#1}}\fi}
\newcommand\msb[1]{\ifmmode
	   \oldmsb{#1}
	 \else \math{\oldmsb{#1}}\fi}

\catcode`@=12

\begin{document}
   \title{High-resolution spectroscopic search for the thermal emission
  of the extrasolar planet HD~217107\,b}

   \author{Patricio E. Cubillos  \inst{1, 2}
          \and
          Patricio Rojo          \inst{1}
          \and
          Jonathan J. Fortney    \inst{3}
          }

   \institute{Department of Astronomy, Universidad de Chile,
  Santiago, Chile
             \\ \email{pcubillos@fulbrightmail.org}
        \and
              Department of Physics, Planetary Sciences Group, University of Central Florida, FL 32817-2385, USA
         \and
             Department of Astronomy and Astrophysics, University of
  California, Santa Cruz,CA 95064, USA\\
             }

\date{Received: - / Accepted: 28 February 2011}
\titlerunning{High Resolution Search of HD~217107\,b}
\authorrunning{Cubillos et al.}

\abstract{We analyzed the combined near-infrared spectrum of a
  star-planet system with thermal emission atmospheric models, based
  on the composition and physical parameters of the system. The main
  objective of this work is to obtain the inclination of the orbit,
  the mass of the exoplanet, and the planet-to-star flux ratio. We
  present the results of our routines on the planetary system
  HD~217107, which was observed with the high-resolution spectrograph
  Phoenix at 2.14 \microns.  We revisited and tuned a correlation
  method to directly search for the high-resolution signature of a
  known non-transiting extrasolar planet. We could not detect the
  planet with our current data, but we present sensitivity estimates
  of our method and the respective constraints on the planetary
  parameters. With a confidence level of 3--$\sigma$ we constrain the
  HD~217107\,b planet-to-star flux ratio to be less than 5\tttt{-3}.
  We also carried out simulations on other planet candidates to assess
  the detectability limit of atmospheric water on realistically
  simulated data sets for this instrument, and we outline an optimized
  observational and selection strategy to increase future
  probabilities of success by considering the optimal observing
  conditions and the most suitable candidates.}
   \maketitle
%

\section{Introduction}
The characterization of the over 500 detected exoplanets has now begun
to take place. Most of the studies are carried out at optical and
infrared wavelengths, because this is where the planetary reflected
light and thermal emission peak, respectively.  The discovery of
transiting planets \citep{Charbonneau00, Henry00} allowed astronomers
to constrain new physical parameters such as the radii and masses of
the planets, which are not measurable by the radial velocity method
alone.  It is on these systems that in the last years the planetary
atmosphere characterization has achieved the most exciting progress
through the use of spectroscopy and broadband photometry with space
telescopes. Examples are the identification of molecules such as water
absorption \citep[e.g.][]{Tinetti07} or methane \citep{Swain08}, or
the observation of the thermal emission variation with orbital phase
\citep{Knutson07}.

Although great improvements in characterizing the composition of
transiting Hot-Jupiters have been achieved, they only represent about
20\% of the known extrasolar planets\footnote{www.exoplanet.eu}.  The
characterization of non transiting planets would require the direct
detection of their light, but the very low flux ratios between the
planets and their host stars makes a direct detection a very
challenging goal.  Secondary eclipse observations from Spitzer show
that planet-to-star flux ratios can be as high as 2.5\tttt{-3} between
3.6 and 24 \microns\ \citep[e.g.][]{Knutson2008}. At 2.14 \microns\
the expected flux should be less than these values.  Many authors have
attempted a direct detection of the Doppler-shifted signature in
high-resolution spectroscopy from ground-based telescopes.  In the
optical \citet{Cameron99} tried to observe the starlight reflected
from the giant exoplanet Tau-Bo\"otis\,b, they found an upper limit to
the albedo and radius using a least-squares deconvolution method that
is well described in the appendices of \citet{Cameron02}, later the
author repeated the analysis on $\upsilon$ Andromeda\,b
\citep{Cameron02}. Recently \citet{Rodler08, Rodler10} searched in the
visible spectra of HD~75289Ab and Tau-Bo\"otis\,b and found upper
limits for their albedos using a model synthesis method. They
constructed a model of the observation composed by a stellar template
plus a shifted and scaled-down version of the stellar template to
simulate the starlight reflected from the planet, these models were
compared to the data by means of $\chi^2$.  In the near-infrared,
several attempts have been made to detect Hot-Jupiters by trying to
distinguish the planetary thermal emission from the starlight
\citep{Wiedemann01, Lucas02, Barnes07, Barnes08, Barnes10}, they also
found upper limits for the emitted flux of the planets.  All these
authors have used their own variation of a method based on the same
principle of separating the planetary and stellar spectra given their
relative Doppler shifts.  Only recently, \citet{Snellen10} claimed the
detection of carbon monoxide from the transmission spectrum of
HD~209458\,b during a transit observation by using high-resolution
spectra; nonetheless, his technique required a transiting system.

In this work we present an effort to constrain new physical parameters
of the non-transiting Hot-Jupiter HD~217107\,b. We attempt to trace
its Doppler-shifted signature (estimated to be \sim 10$^{-4}$ times
dimmer than the star flux) with a correlation function between
high-resolution data and models of its atmospheric spectrum. With
positive detections this method would provide new information on its
characteristics, such as its temperature, chemical composition, and
the presence of chemical tracers associated with life. At the same
time, the method enables the calibration of high-resolution
spectroscopic models for a larger sample of planets that do not
necessarily transit their parent star.

In Section \ref{HD217} we review the planetary system HD~217107; in
Section \ref{data} we describe the observations, data reduction, and
calibration procedures; in Section \ref{analysis} we detail the
theoretical atmospheric spectrum of the planet and the method used to
extract and analyze the planetary signal and present the results of
our data; in Section \ref{simulations} we develop a strategy for the
ideal data acquisition situation and simulate observations of other
planetary systems; and in Section \ref{discussion} we give the
conclusions of our work.

\section{The planetary system HD~217107}
\label{HD217}

\subsection{HD~217107\,b discovery}
\label{HD217107b}

HD~217107 is a main-sequence star that is similar to the Sun in mass,
radius, and effective temperature; its spectral type, G8 IV, indicates
that it is starting to evolve into the red-giant phase
\citep{Wittenmyer07}.  The presence of HD~217107\,b was first reported
by \citet{Fischer99} through radial velocity measurements of the star,
the detection was then confirmed by \citet{Naef01}. Later,
\citet{Fischer01} identified a trend in the residuals of the fit, and
\citet{Vogt05} postulated the existence of a third companion in an
external orbit with a period of 8.6 \pm\ 2.7 yr. The presence of this
third object promoted the study of this system in subsequent surveys
\citep{Butler06, Wittenmyer07, Wright09}, constraining more precisely
the companions' orbital parameters.  Table \ref{tabpars} summarizes
the parameters used in this work.

\begin{table}[h]
\centering
\caption{Orbital parameters of HD 217107. \label{tabpars}}
\begin{tabular}{lr@{\,\pm\,}l@{\,}c}
\hline\hline
Parameter &        \mctc{Value} & References\tablefootmark{a} \\
\hline
Star: \\
Spectral type                      & \mctc{G8 IV}               & W07 \\
\math{T_\mathrm{eff}}(K)           & 5\,646     & 26            & W07 \\
K (mag)                            & 4.536      & 0.021         & C03 \\
\math{d} (pc)                      & 19.72      & 0.30          & P97 \\ 
\math{M\sb{s}} (M\math{\sb{\odot}}) & 1.02      & 0.05          & S04 \\
\math{K\sb{s}} (m\,s\math{\sp{-1}}) & 140.6     & 0.7           & W07 \\
\math{v\sb{g}} (km\,s\math{\sp{-1}})& -14.0     & 0.6           & N04 \\
\hline
Planet: \\
\math{P} (days)                    & 7.12689    & 0.00005       & W07 \\ 
\math{T\sb{p}} (JD)                & 2\,449\,998.50 & 0.04      & W07 \\ 
\math{e}                           & 0.132      & 0.005         & W07 \\ 
\math{m\sb{p} \sin i} (M\sb{\rm Jup}) & 1.33    & 0.05          & W07 \\ 
\math{a} (AU)                      &  0.074     & 0.001         & W07 \\ 
\math{\omega} (deg)                & 22.7       & 2.0           & W07 \\
\hline
\end{tabular}

\tablefoottext{a}{W07: \citet{Wittenmyer07}, C03: \citet{Cutri03}, \\
  P97: \citet{Perryman97}, S04: \citet{Santos04}.}
\end{table}

\subsection{ Radial velocity}
\label{subsecpc}

The radial velocity of the planet, \math{v\sb{p}\sin i}, around the
center of mass of the system is given by the reflex motion of the
star:

\begin{equation}
\label{eq:twobody}
v_p(t)\sin i\ =\ -\ v_s(t)\sin i\ \frac{m_s}{m_p\sin i}\times\sin i .
\end{equation}

\begin{figure}[t]
\centering
\includegraphics[trim=15 0 50 0, width=\columnwidth, clip]{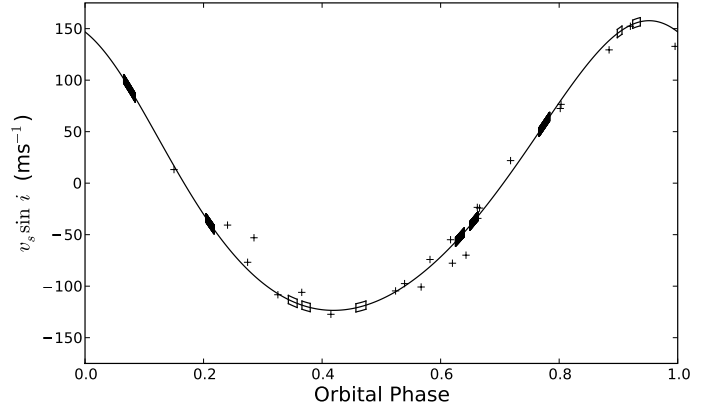}
\caption{ Radial velocity curve of HD 217107 \vs\ orbital phase. The
  crosses mark the observations of \citet{Wittenmyer07}, which we used to
  compute this orbital solution. 
  The boxes over the curve indicate the coverage of our observations, the
  filled boxes represent the runs utilized in the analysis, while the
  open boxes represent the discarded runs (details in Section
  \ref{results}). }
\label{rvfig}
\end{figure}

It depends on the mass of the star, $m\sb{s}$; the minimum mass of the
planet, $m\sb{p}\sin i$; the projected radial velocity curve of the
host star, $v\sb{s}(t)\sin i$ (which in turn depends on the parameters
$T\sb{p},\ P,\ e,\ \omega,\ {\rm and}\ K\sb{s}$); and the inclination
of the orbit, $i$, and also on the velocity of the center of mass of
the system, $v\sb{g}$, when measured from Earth.  Thus, the radial
velocity curve of the planet is a distinctive curve in time,
parameterized by the values summarized in Table \ref{tabpars}, where
the only unknown parameter is the inclination of the orbit.  Figure
\ref{rvfig} shows the radial velocity curve of the star owing to the
interaction with HD~217107\,b, phased over one orbit, with the origin
in phase ($\phi = 0$) at the time of periastron. The radial velocity
of the planet is proportional to this radial velocity curve (Equation
\ref{eq:twobody}).

\subsection{ Flux estimate}
\label{estimate}

By simulating the spectra of the planet and its host star as black
bodies, we can estimate the order of magnitude of the planet-to-star
flux ratio as a function of wavelength.  The black body emission,
$F_\lambda(T)$, is determined by the surface temperature of the
object. While for the star the temperature is well known from models
(see Table \ref{tabpars}), for the planet our best approximation is
the equilibrium temperature
\begin{equation}
T_{eq}  =  \left( \frac{1-A}{4}   \right)^{1/4}
           \left( \frac{R_{s}}{a} \right)^{1/2} T_\mathrm{eff}.
\label{eq:teq}
\end{equation}

\begin{figure}[h]
\centering
\includegraphics[trim=10 0 15 0, width=\columnwidth, clip]{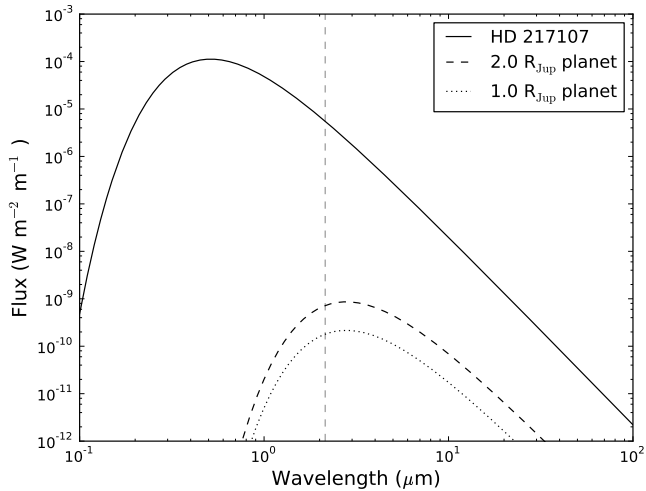}
\caption{Black body emission of HD~217107 and  HD~217107\,b
  assuming a planet radius of 1.0 and 2.0 Jupiter
  radii. The vertical dashed line marks the waveband of our data
  (2.14 \microns). The directly reflected light component has
  little contribution in the infrared and is thus omitted.}
\label{fig:bbrad}
\end{figure}

For a reference value of the bond albedo of $A=0$, we found an
equilibrium temperature for HD~217107\,b of $T_{eq} = 1040 \pm
19$~K. Figure \ref{fig:bbrad} shows the black body spectrum of the
star and the planet assuming a radius between one and two Jupiter
radii, which is the range of the radii for giant extrasolar planets
measured to date.

The planet-to-star flux ratio is given by
\begin{equation}
{\rm Flux\ ratio} \ =\ \frac{F_\lambda(T_{planet})}{F_\lambda(T_{star})}
                    =\ \frac{B_\lambda(T=1040\ \rm K)}{B_\lambda(T=5646\ \rm K)} 
                       \left(\frac{R_p}{R_s}\right)^2 .
\end{equation}

At 2.14 \microns, the flux ratio varies between 3\tttt{-5} and
1.5\tttt{-4} from one to two Jupiter radii of the planet's radii,
respectively.  For shorter wavelengths the flux ratio decreases,
because the star light dominates the emission spectrum. For longer
wavelengths the net fluxes and thus the signal to noise ratio are
lower.

\section{Observations and data reduction}
\label{data}

\subsection{ Observations }
\label{observations}

We observed the planetary system HD~217107 in 11 nights between 2007
August 14 and November 28 using Phoenix \citep{Hinkle03}, a
high-resolution near-infrared spectrometer at the Gemini South
Observatory.

The spectrograph has a 256 x 1024 InSb Aladdin II array with a
resolving power of \math{\ttt{-5}} \microns\ per pixel, the slit
covers 14 arc seconds in length. Its gain is 9.2~e\math{\sp{-}}/ADU
and it has a readout noise of 40~electrons. An argon hollow cathode
wavelength calibration source is supplied with the instrument.  Over
950 frames of the system were obtained in service mode, using the
standard ABBA nodding sequences to easily remove sky emission, they
cover a portion in the infrared spectral range from 2.136 to 2.145
\microns\ (see Table \ref{tab:obs}).  We tuned the data acquisition
after receiving the data from the first runs since the instrument was
not fully characterized for use on the Gemini Telescope.  For the
first two nights, the exposure time was set to 25 seconds, whereas for
the rest of the nights it was set to 80 seconds. We requested arc-lamp
calibration exposures as well.

\begin{table}[h]
\caption{Phoenix observations of HD 217107. \label{tab:obs}}
\centering
\begin{tabular}{ccccc}
\hline\hline

Date  & Time on Target\tablefootmark{a}  & Orbital Phase  & 
  $\Delta v$\tablefootmark{b}  & Status\tablefootmark{c} \\
UT  & min  &  & ms\sp{-1}  & \\
\hline
2007-08-14 & \phantom{0}45  & 0.30 &           15.72 &          \\ 
2007-08-16 & \phantom{0}45  & 0.61 &           12.29 &          \\ 
2007-08-22 & \phantom{0}22  & 0.43 & \phantom{0}2.77 & rejected \\ 
2007-08-26 & \phantom{0}45  & 0.99 & \phantom{0}3.57 & rejected \\ 
2007-10-02 &           192  & 0.16 &           25.51 &          \\ 
2007-11-19 & \phantom{0}96  & 0.90 & \phantom{0}2.61 & rejected \\ 
2007-11-23 & \phantom{0}96  & 0.46 & \phantom{0}2.79 & rejected \\ 
2007-11-24 & \phantom{0}96  & 0.60 &           13.90 &          \\ 
2007-11-25 & \phantom{0}96  & 0.74 &           12.33 &          \\ 
2007-11-26 & \phantom{0}96  & 0.88 & \phantom{0}3.84 & rejected \\ 
2007-11-28 & \phantom{0}96  & 0.16 &           10.23 &          \\ 
\hline 
\end{tabular}

\tablefoottext{a}{Total exposure time of HD~217107.\\}
\tablefoottext{b}{Radial velocity span of the star during the observing time.\\}
\tablefoottext{c}{See Sections \ref{subseccc} and \ref{results} for details.\\}

\end{table}

\subsection{Reduction}
\label{reduction}

We wrote our own interactive data language
(IDL)\footnote{http://www.ittvis.com/ProductServices/IDL.aspx}
routines for the data reduction and analysis, processing each night
and slit position as an independent data set to minimize systematics
caused by different atmospheric conditions or instrumental set-up. We
used the flat-field images to identify hot pixels, marking a pixel as
bad if it had a value beyond 3.5 sigma from the median of the values
of the nine subsequent pixels in its neighborhood. Bad pixels were
masked in all further processing stages. Then, we divided the frames
by a per-night master flat-field and subtracted their corresponding
opposite A or B frame to remove bias and sky. Finally, we extracted
the spectra from the frames with an IDL
implementation\footnote{http://physics.ucf.edu/\math{\sim}jh/ast/software/optspecextr-0.3.1/doc/index.html}
of the optimal spectrum extraction algorithm described in
\citet{Horne86}, this algorithm identified cosmic ray hits, which were
also masked from subsequent processing.

\subsection{Wavelength calibration}
\label{subsecwc}

First, we calibrated the wavelength dispersion using the ThAr lamps,
identifying the line positions and strengths in a high-resolution ThAr
line atlas \citep{Hinkle01}.  Because there was only one calibration
lamp for each night, this solution represented only a rough wavelength
calibration, because there are (sub pixel) offsets in wavelength in
the data.  To reach the high precision needed for this work, we
fine-tuned the calibration with a high-resolution spectrum of the
Sun\footnote{http://bass2000.obspm.fr/solar\_spect.php} to identify
the telluric lines (identified as those present both in the solar
spectrum and in an average spectrum of our data set).

We constructed an average spectrum to increase the S/N ratio by
aligning and adding the spectra of each night. To determine the
relative shifts, we selected within each set the first spectrum as
reference, while the rest were shifted (using spline interpolations)
to calculate the shift that minimized the root-mean-square of the
correlation with the reference.  The centers of fifteen common
absorption lines were identified in wavelength values for the solar
spectrum and in pixel position for our average spectrum. The
wavelength solution is obtained by fitting a second-order polynomial
($\lambda = c_0 + c_1\cdot p + c_2\cdot p^2$) to the solar wavelength
\vs\ the pixel position.  Typical fitting coefficients are $c_0=
2.145407$, $c_1=-1.0305\tttt{-5}$, and $c_2=-1.8496\tttt{-10}$. The
dispersion of the residuals is $RMS=4.21\tttt{-6}$ \microns.  No
pattern is seen in the residuals.  Pixels at wavelengths dominated by
the identified telluric absorption lines were discarded from
subsequent processing owing to their highly variable nature. About
65\% of the pixels remained for the next analysis steps.

\section{ Data analysis and results}
\label{analysis}

\subsection{ Correlation }
\label{subseccc}
Because it is impossible to directly distinguish the planet's
signature from the stellar one in a single spectrum, following the
idea of \citet{Deming00} and \citet{Wiedemann01}, we searched for the
planetary Doppler-shift signature through a correlation method between
the (stellar-subtracted) residual data and a synthetic model of the
planet's spectrum.  To remove the stellar flux, we aligned the spectra
for each set (Doppler-shifting them and using a spline interpolation)
in a reference system in which the star remains at rest, and
constructed a stellar template from the average of the set.  Then, the
stellar templates and the spectra are normalized dividing by their
respective medians. Finally, the wavelengths of the stellar templates
are shifted according to the orbital phase of the star in each
individual spectrum, and then the stellar template is subtracted from
them.  We avoided combining the different nights to obtain the stellar
template, because it is highly probable that other systematics would
be introduced.

\begin{figure}[t]
\centering
\includegraphics[trim=30 0 55 0, width=\columnwidth, clip]{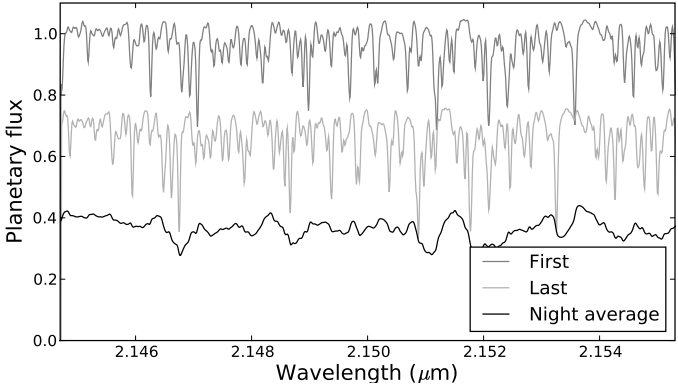}
\caption{Planetary spectrum blurring in the stellar template. Using
  our synthetic spectra of HD~217107\,b we simulated the smearing of
  the planetary spectrum over one observing run. The dark-gray and
  light-gray lines denote the first and last spectra of HD~217107\,b
  during an observing night. The relative shift owing to the orbital
  motion of HD 217107 b is 40 kms$^{-1}$ in this simulation. The
  bottom black line shows averaged the planetary spectra.}
\label{smoothing}
\end{figure}

Because the planet is approximately a thousand times less massive than
its host star, the planetary Doppler wobble is greater by the same
order of magnitude (see Eq. \ref{eq:twobody}), consequently the
planetary signature will not be added coherently in the stellar
template and thus appear blurred. The stellar template subtraction
leaves a residual spectrum that consists of the signature of the
planet, which is slightly attenuated in the averaging process and
immersed in Poisson noise.  The blurring of the planet signature (see
Figure \ref{smoothing}) is determined by the planetary velocity span,
which in turn depends on the time span of an observation and the
orbital phase at the time of the observation. Observations near
inferior or superior conjunction provide the greatest radial velocity
spans, while observations close to the greater elongation of the
planet's orbit produce the smallest radial velocity spans, rendering
the data useless.  The rejected data sets in Table \ref{tab:obs} were
observed near greater elongation.

For the high-resolution synthetic planetary spectra of HD~217107\,b we
used customized theoretical thermal emission models of its atmosphere
\citep[model described in][]{Fortney05, Fortney06, Fortney08} at three
different distances from the star to account for the non-negligible
eccentricity of the planet.  The models are cloud-free, with solar
metallicity, gravity $g=20$~m\,s$^{-2}$, and the molecular abundances
are those appropriate for chemical equilibrium.  At these effective
temperatures, the main absorbing molecules are H$_2$O, CH$_4$, CO, and
CO$_2$.  The chemistry is described in detail in \citet{Lodders02} and
\citet{Visscher06}.  We empirically characterized the instrumental
resolution through the analysis of the emission lines in the
calibration lamps.  We convolved the model spectra by the instrumental
resolution, which we determined to be \math{\lambda/\Delta\lambda
  \approx 40\,000}.  Then, for an assumed value of \math{\sin i}, the
synthetic spectra are Doppler-shifted to mimic the radial velocity of
the planet at the time that the data frame was obtained. The
correlation degree, $C(i)$, is calculated according to the formula

\begin{equation}
C(i)  =  \frac{1}{N}\sum\sb{k=1}\sp{N} \frac{ \sum\sb{j=1}\sp{N\sb{k}} 
           \left(  r\sb{kj}      - \bar r\sb{k}       \right) \cdot
           \left( \tau\sb{kj}(i) - \bar \tau_k(i) \right)   }{ \sqrt{
 \left\{ \sum\sb{j=1}\sp{N_k} \left( r\sb{kj} - \bar r_k \right)^2 \right\}
 \left\{ \sum\sb{j=1}\sp{N_k} \left( \tau\sb{kj}(i) - \bar \tau_k(i) \right)^2
\right\}       } } .
\label{cross}
\end{equation}

In this equation we used the notation $f\sb{kj}$ for the value of
the function at the pixel $j$ of the spectrum $k$, and $\bar f\sb{k}$
for the mean value of the function in the spectrum $k$.  Here, ``$r$''
refers to the residual spectrum while ``$\tau(i)$'' to the shifted
planetary model spectrum, with $N_k$ the number of pixels in spectrum
$k$ and $N$ the total number of spectra.  The denominator in the
expression normalizes the correlation, and thus a value of 1.0 would
indicate a perfect correlation.

We thus produce a curve of the correlation degree \vs\ the inclination
of the orbit, evaluated in the range $0 < i < \pi/2$.  A positive
value of this function indicates that the data spectrum resembles that
of the model, while a negative one suggests anti-correlation.  As
consequence of the random nature of the Poisson noise, the value of
the correlation between the residual spectra and the models should be
close to zero, except when the adopted $i$ matches that of the
planetary system.  Therefore, an appreciable peak in the correlation
curve would represent a successful detection of the planetary
signature and immediately indicates the value of $i$.  By constraining
the inclination with this method, the mass of the planet would be
immediately determined via Eq. 1.

\subsection{Data results}
\label{results}

For this analysis, we excluded the nights where the velocity span of
the star was less than 10 m\,s\sp{-1} (column 4 of Table
\ref{tab:obs}) since they do not represent any significant improvement
in the results, because the shift of the planet ($\sim8.3$
km\,s$^{-1}$) is not significantly higher than the instrumental
resolution ($\sim7.5$ km\,s$^{-1}$).  Figure \ref{fig:results} (Top
panel) shows the correlation curve derived from our data as a function
of $\sin i$.  The degree of correlation found was close to zero at all
inclinations, and we do not distinguish any identifiable positive peak
that could indicate an atmosphere with absorption features resembling
those of the models.

\subsection{Planet-to-star flux ratio fitting} 
\label{detlim}

\begin{figure}[t]
\centering
\includegraphics[trim = 5 5 15 5, width=\columnwidth, clip]{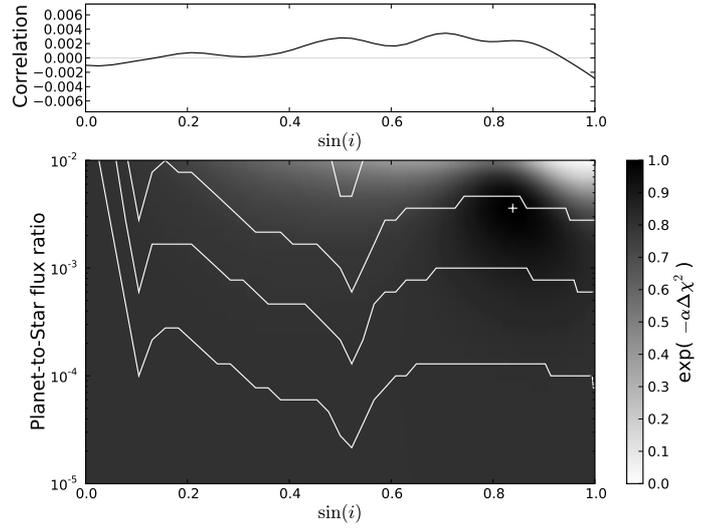}
\caption{ Top: Correlation result for our data set as a function of
  \math{\sin i}.  The correlation remains flat along every inclination
  without any distinctive peak, the maximum value is reached at
  \math{\sin i=0.71}.  Bottom: Goodness of fit, \math{\chi\sp{2}}-map,
  of the correlation models to the data. The horizontal and vertical
  axes refer to the fitting parameters \math{\sin i} and
  \math{F\sb{p}/F\sb{s}} respectively, at which the synthetic
  planetary spectrum was added in the correlation models creation,
  from which we calculated the minimum squares ($\chi^2\sb{i,{\rm
      fr}}$). We plot $\chi^2$ relative to the best fit ($\Delta
  \chi\sp{2}\sb{i,{\rm fr}} = \chi\sp{2}\sb{i,{\rm fr}} -
  \chi\sp{2}\sb{{\rm min}}$) using the function $\exp(-\alpha\cdot
  \Delta \chi\sp{2})$. The gray scale denotes the goodness of fit,
  from black for the best fit (at $\chi\sp{2}\sb{{\rm min}}$), to
  white for the poorest fit.  The plotting parameter, \math{\alpha},
  just enables a good contrast in the plots (the same value was used
  for all plots).  Additionally, we determined with bootstrap
  procedures the solid lines (bottom to top) that mark the (1, 2, 3
  and 4--\math{\sigma}) levels of false-alarm probability. The white
  cross marks the best fit at \math{\sin i = 0.84} and
  \math{F\sb{p}/F\sb{s} = 3.6 \tttt{-3}}, situated below the
  3--\math{\sigma} confidence level.}
\label{fig:results}
\end{figure}

While the inclination determines the maximum of the correlation curve,
the planet-to-star flux ratio (\psfr) is the main physical parameter
bounded to the magnitude of the correlation.  In this section we
determine the most probable values in the parameter space [\psfr\,,\
$\sin i$], which gave rise to our result, and estimate the statistical
significance of the value of the correlation reached.  We searched for
the best fitting values comparing our data results
(Fig. \ref{fig:results} Top) with ``synthetic'' correlation curves.
We generated the synthetic correlation curves by recreating our
observations, adding a synthetic planetary spectrum, with known
inclination and planet-to-star flux ratio, according to the following
scheme:
\\
Step 1: We rearranged the order of the data set with random
permutations within each night, but kept the original order of the
dates of the observations.  As a consequence, any real planet
signature disappeared, but the noise level of the data was conserved.
\\
Step 2: Using the atmospheric models of the planet, we injected a
synthetic spectrum in the scrambled data set, Doppler-shifted and with
a relative flux according to specific values of $\sin i$ and \psfr,
respectively. For simplicity, we adopted a constant \psfr\ along the
orbit.
\\
Step 3: We processed these synthetic data through the same routines as
in our original data (section \ref{subseccc}). We then iterated for a
grid of values in the ranges: $0 \le i \le \pi/2$ and $\ttt{-5} \le
\psfr \le \ttt{-2}$, obtaining a set of synthetic correlations for
$\sin i$ and \psfr.

Once we obtained these models, we searched for the best-fit parameters
through a $\chi^2$ minimization between the data correlation curve
and the synthetic correlation curves, generating a goodness-of-fit
map (Fig. \ref{fig:results}, bottom panel).

In addition, we used a bootstrap procedure to calculate
false-alarm-probability limits for this map.  Following
\citet{Cameron02}, we determined the frequency with which the
correlation degree exceeds a given value as a result of noise in the
absence of a planet signal.  The routine consists of performing a
random permutation of the data sets and the subsequent data analysis
(steps 1 and 3 of previous paragraphs) which we repeated a large
number of times (\sim 5\,000), recording the correlation curve after
each trial.  This set of correlation curves represents the correlation
found in the absence of a planetary signal, and, because it is created
from the data themselves, defines an empirical probability
distribution that includes both the photon statistics and instrumental
systematics.

Then, at each inclination, we stacked and sorted the values of the
correlation in increasing order. We determined the 1, 2, 3, and
4--\math{\sigma} false-alarm confidence levels as the value of the
correlation degree at the 65, 90, 99 and 99.9 percentiles of the
trials.  They represent the signal strengths at which spurious
detections occur with 35, 10, 1, and 0.1 percent false-alarm
probability respectively, at each value of the inclination.  This
allows us to assess the probability of obtaining a certain correlation
degree in the absence of planetary emission.

Fig. \ref{fig:results} (Bottom panel) shows the probability map for
HD~217107\,b. The best fit occurs at $\sin i = 0.838$ and
$F\sb{p}/F\sb{s}=3.6\tttt{-3}$, although the relative improvement in
$\chi^2$ against the surrounding parameters is shallow.  This value
disagrees with the maximum value of the correlation curve
(Fig. \ref{fig:results} Top), and furthermore, the bootstrap results
indicate that this value is below the 3--$\sigma$ confidence limit of
the signal not being a false positive. Also, this \psfr\ is much
higher than the predicted value from Sec. \ref{estimate} (between
3\tttt{-5} and 1.5\tttt{-4}).  The disagreement of the results of the
top and bottom panel in Fig. \ref{fig:results} suggests that
systematics remain after the data reduction, while the strength of the
correlation value, two orders of magnitude above the expected flux
ratio, indicates that this result is not realistic.  The 3--$\sigma$
confidence limit only allows us to establish an upper limit in the
flux ratio at 4--5 \tttt{-3} for inclinations greater than $\sin i =
0.6$.  In conclusion, we cannot state the detection of \twoseventeenb.

\section{Future prospects}
\label{simulations}

\subsection{Observational strategy}\label{obstrat}

Although our current data do not enable us to claim the detection of
HD~217107\,b, we identified a strategy to maximize the chances of a
successful detection. This involves selecting suitable candidate
systems and precisely choosing the phasing and span of the
observations.  To exemplify the advantages, we simulated realistic
observations of other planetary systems.

\begin{figure*}[t]
\includegraphics[trim = 5 5 13 5, scale=0.425, clip]{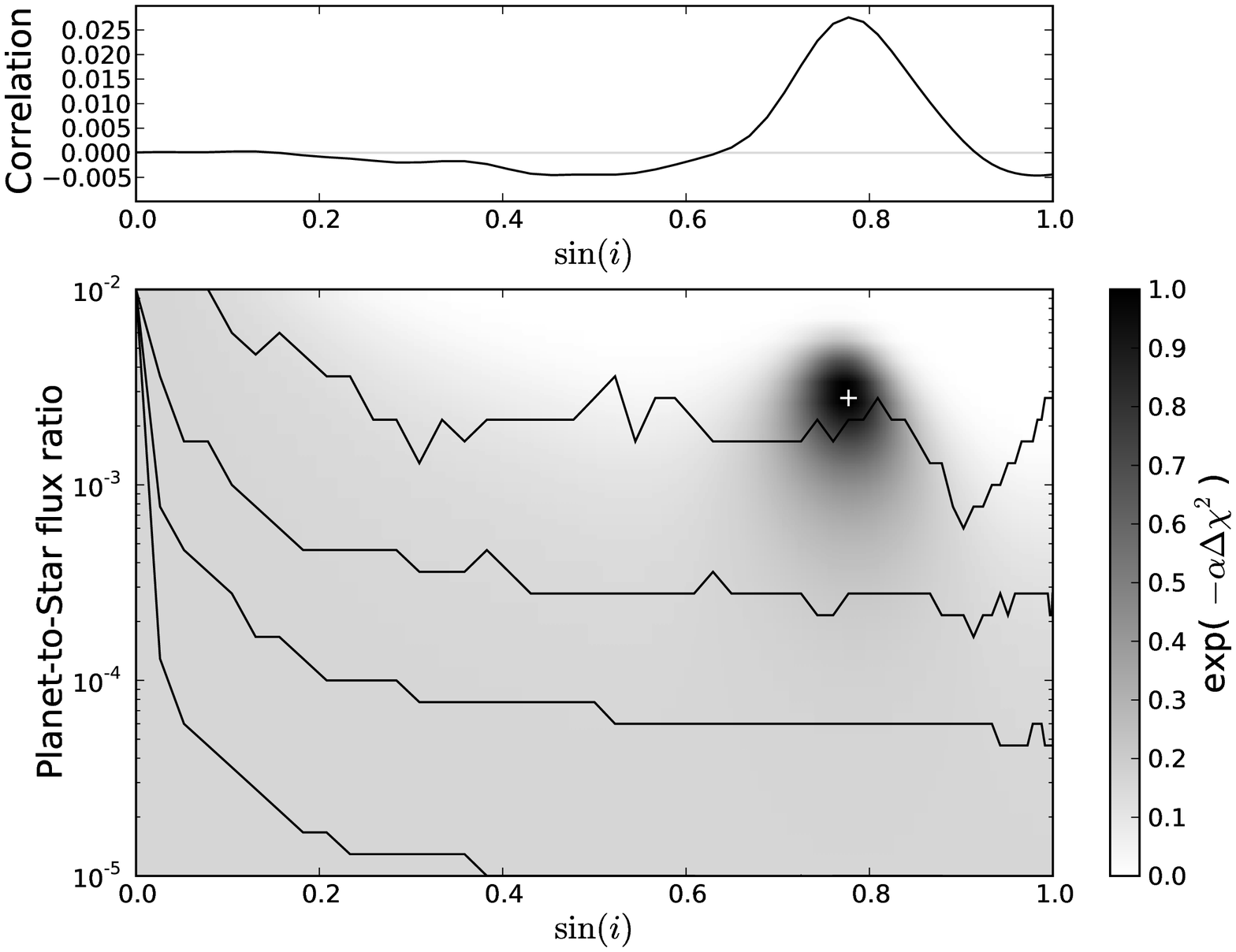}
\includegraphics[trim = 7 5 15 5, scale=0.425, clip]{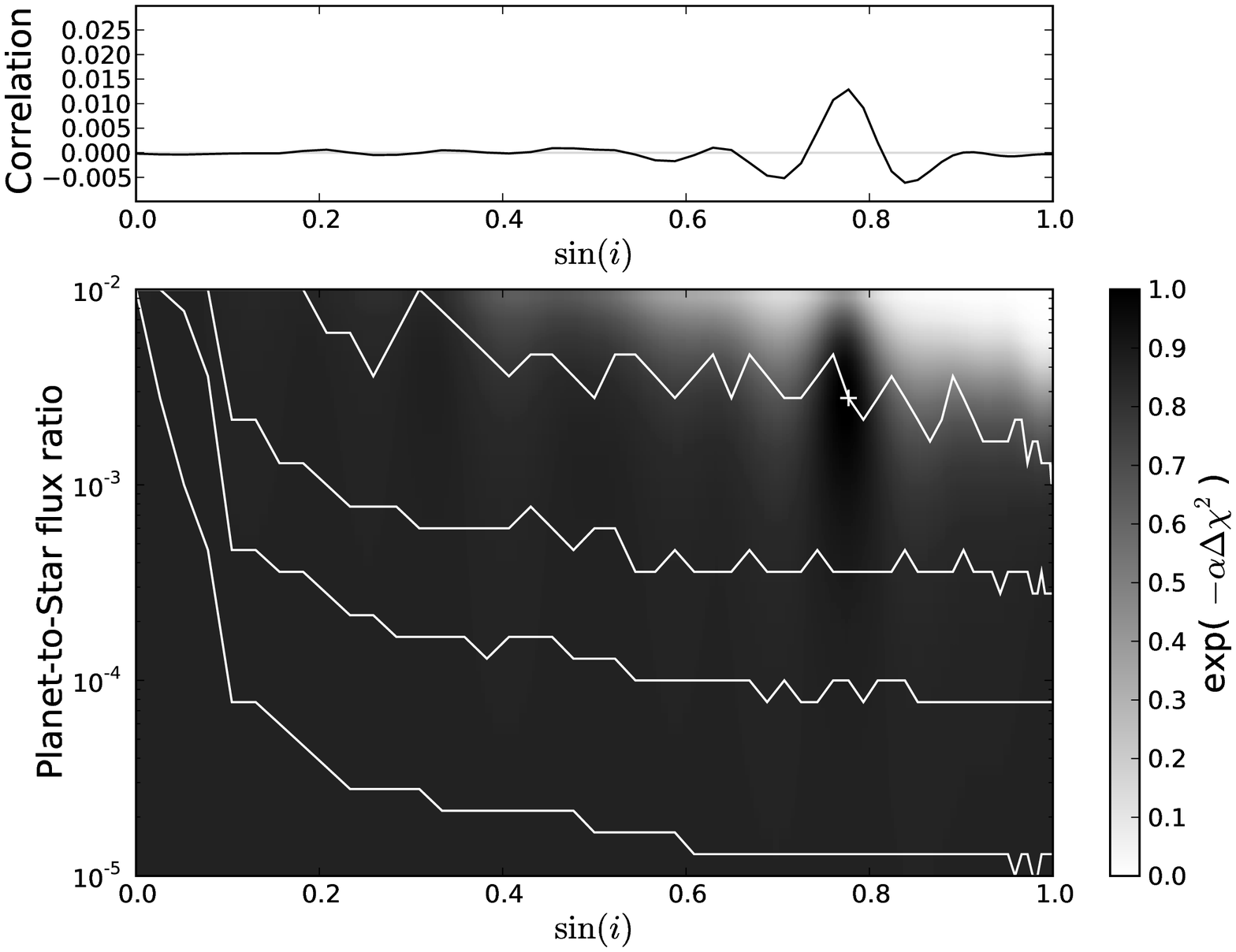}
\caption{ Correlation curves and \math{\chi\sp{2}}-maps of synthetic
  data of HD 179949. A synthetic planetary signal was injected in the
  spectra with 
  parameters: \math{\sin i=0.77} and \math{F\sb{p}/F\sb{s}=0.003}.
  Left: Using our observing strategy.  Right: random distributed
  observing dates. The routine successfully recovers the signal at
  \math{\sin i = 0.78} and \math{F\sb{p}/F\sb{s} = 2.8 \tttt{-3}} in
  both cases, although, when using our strategy, the correlation
  degree is stronger, and the parameters are better determined
  compared with the right panel.}
\label{sim_179}
\end{figure*}

We limited our sample to the currently known extrasolar planets
without transits\footnote{http://exoplanet.eu}, observables from
Gemini South, for the first semester of the year.  Even though we
constrained ourselves to the instrument we characterized and to a
fixed time span, the purpose of these simulations is to provide one
successful detection with our method. It is plausible that considering
the full extent of possibilities, stronger signals can be acquired.
The improvement in a detection, limited by purely photon noise, can be
quantified by the planet-to-star flux ratio and the stellar flux,
according to the expression (fluxes in number of photons)
\begin{equation}
\label{eq:snr}
F_p/{\rm Noise}\ =\ \frac{(\psfr) \cdot F_s}{\sqrt{|F_s|+|F_p|}}\ 
           \approx\ (\psfr)\cdot\sqrt F_s .
\end{equation}
Then, for example, the spectrometer CRIRES with four times the
wavelength coverage of that of Phoenix, has twice the sensitivity of
Phoenix.  We decided to simulate the Phoenix instrument, since it is
well characterized by our group, while other instruments should
present their own systematics, which are hard to quantify.

We simulated the planets as if the strength of the high-resolution
absorption features were the same as that of our models for
HD~217107\,b, but with the corresponding \psfr\ (estimated as in
Sec. \ref{estimate}). A caveat for this assumption is that the
strength of the lines is not very clear in planets that exhibit
thermal inversions. \citet{Burrows2008} and \citet{Fortney08} suggest
that the emission features should be weaker.

The target selection criteria are based first on the radial velocity
span of the planet, where we set a lower limit cutoff of 7.5
km\,s\sp{-1} (equivalent to the FWHM of the instrument spectral
resolution) for a three-hour observing run if the orbit was at \math{
  i = 30\sp{\circ}}.  Second, we look for higher apparent brightnesses
of the stars for better signal-to-noise ratios. Table \ref{tabplanets}
lists two of the better suited selected planetary systems (HD~217107
listed for comparison). A brighter K-band magnitude of the star
improves the signal-to-noise ratio, while a smaller semi-major axis
involves a higher radial velocity span, which enables a greater
Doppler shift of the planet spectra during the runs and at the same
time favors higher planet-to-star flux ratios.

\begin{table}[b]
\caption{ Favorable targets for Gemini South. \label{tabplanets} }
\centering
\begin{tabular}{lcccc}
\hline\hline
Target & \math{a} & M\sb{K} & \math{K\sb{p}} \\
       & AU       &         & km\,s\sp{-1} \\
\hline
HD~179949    & 0.045   & 4.94    & 158.23   \\ 
Tau Boo      & 0.046   & 3.51    & 150.62   \\ 
HD~217107    & 0.073   & 4.54    & 112.28   \\  
\hline 
\end{tabular}
\tablefoot{ Planet's radial velocity amplitude for an
 orbit with \math{\sin i=1}.}
\end{table}

To simulate the observations, we recreated the same instrumental
settings of our data, but carefully selected the observing schedule.
For each one of the nights in the period and restricted to air masses
under 1.5, we selected the three-hour range that gives the maximum
velocity span.  We recorded then, the radial velocity spans for each
night, and chose those with the biggest spans.  We used the solar
spectrum to simulate the stellar spectrum, while for the planetary
component we used the atmospheric models of HD~217107\,b added with a
given planet-to-star flux ratio and inclination. Each component is
Doppler-shifted according to the orbital parameters. Finally we added
Poisson noise to the spectra, according to the signal-to-noise
corresponding to the magnitude of the target. The synthetic data were
processed in exactly the same way as our original data.

\subsection{ Simulations }
\label{simu}

In our first test, we present two simulations of an observing campaign
on a target with the physical parameters of HD~179949 to show the
improvements of our observing strategy in contrast with a regular
observation. The given parameters are \math{\sin i = 0.77} and
\math{F\sb{p}/F\sb{s} = 3 \tttt{-3}}.  Figure \ref{sim_179} left shows
the simulation following our observing strategy, while Fig.
\ref{sim_179} right shows the simulation selecting random observing
dates.  In both cases the correlation curves (top panels) mark the
inclination of the synthetic orbit with a increment in the correlation
degree near $\sin i = 0.77$, while the $\chi^2$-maps (bottom panels)
effectively indicate the best fit at $\sin i = 0.78$ and
$F\sb{p}/F\sb{s} = 2.8 \tttt{-3}$.

We identify the main differences between these two simulations: First,
given the larger radial velocity spans when implementing our strategy,
the planetary spectrum is more blurred in the stellar template and
consequently less reduced when the template is subtracted, the
planetary spectrum signal is thus stronger in the residual spectrum,
which increases the correlation degree.  As consequence of these
greater correlation degrees, all confidence levels are generally
lower, because it is less probable to reach this correlation degree by
chance in the no-planet case, and lastly, the $\chi^2$-map peak is
much better determined.  The improvement is reflected more in the
distinction of the best fit against other values of the parameter
space than in the distinction against the no-planet case.

\begin{figure}[t]
\includegraphics[trim = 5 5 13 5, scale=0.425, clip]{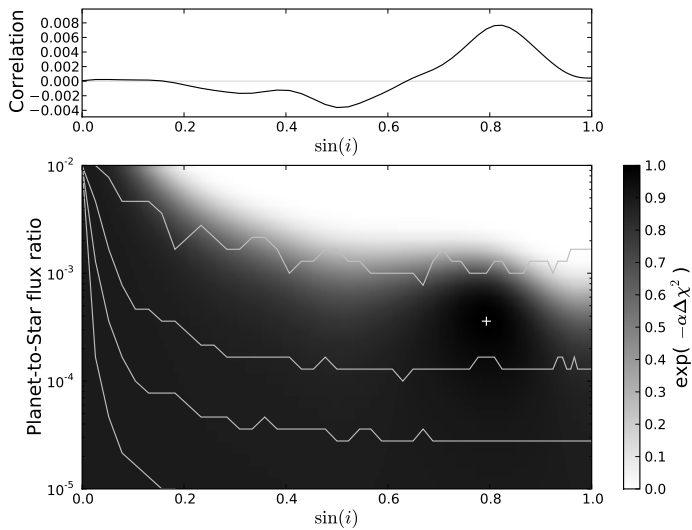}
\caption{\label{sim_others} Similar to Fig. \ref{fig:results},
  correlation curve (top panel) and \math{\chi\sp{2}}-map (bottom
  panel) of a simulation of the planetary system Tau~Boo, with an
  injected companion at \math{\sin i = 0.82} and \math{F\sb{p}/F\sb{s}
    = 4 \tttt{-4}}.}
\end{figure}

In another simulation, we recreated the planetary system Tau Boo as
close as possible to its real physical characteristics
(Figure \ref{sim_others}). Tau~Boo\,b had the parameters $\sin i = 0.82$
and $F\sb{p}/F\sb{s}=4\tttt{-4}$, our routines returned the best fit:
\math{\sin i = 0.79} and \math{F\sb{p}/F\sb{s}=3.6\tttt{-4}},
slightly underestimating the values. Nevertheless, the
\math{\chi\sp{2}}-map shows an improvement in the region near the
injected inclination and flux ratio. The bootstrap results set the
3--$\sigma$ confidence limit near $F\sb{p}/F\sb{s}$ 1.5 \tttt{-4} (for
$\sin i > 0.5$), indicating a detection with 99\% confidence.

\section{Discussion and conclusions}
\label{discussion}

Because the instrument was not well characterized at the time and our
service-mode observational strategy had to be adapted after the first
few observing windows, the data for HD~217107 were not as sensitive as
expected.  The correlation curve was featureless for all inclinations
and with values close to zero, with a maximum at $\sin i = 0.71$.  By
fitting the sine of the inclination and the planet-to-star flux ratio
through least-squares, we found the best-fit parameters of $\sin i =
0.84$ and $F\sb{p}/F\sb{s} = 3.6 \tttt{-3}$ at a level below our
3--\math{\sigma} confident limit.  As a consequence of the faint
features in the results, the disagreement between the peak in the
correlation (Fig. \ref{fig:results} Top panel) and the most probable
value of \math{\sin i} (Fig. \ref{fig:results} Bottom panel), and the
higher than predicted $F\sb{p}/F\sb{s}$, we could not claim a
detection of HD~217107\,b with our current data. Given the results of
the bootstrap procedure, we reject the flux ratio of HD~217107\,b to
its host star to be over 5 \tttt{-3} (3--\math{\sigma} confidence).
We attribute these results to the absence of an ideal strategy in the
data acquisition at the time of the observations and a needed further
treatment of the instrument systematics.

We could not detect HD~217107\,b, but defined the outlines of future
campaigns by carefully defining a candidate selection criterion and an
observational strategy.  We conclude that the best-suited candidates
for this technique are those in very close orbits, which allow the
planets to have high orbital velocities and higher planet-to-star flux
ratios.  We propose an observing strategy where for the period of
observations we specifically select the nights with maximum radial
velocity spans.  To explore the capabilities of our routines, we
simulated other planetary systems as observed by the Phoenix
spectrograph, with the same number of hours and an appropriate
schedule of observations.  The system HD~179949 was recreated,
contrasting the use of our observing strategy with a regular
observation schedule. We recovered the planetary signature in both
cases, but showing an improvement in the correlation degree, precision
in the \math{\chi\sp{2}}-map, and lower \math{\sigma} limits when
using our observing schedule.  Finally, we performed a realistic
simulation of the planetary system Tau Boo, and successfully detected
its signature.

In conclusion high-resolution instruments like Phoenix are capable of
detecting extrasolar planet Doppler-shifted signals with flux ratios
as low as \ttt{4} with this method if we perform a careful treatment
of the systematics (approaching the photon noise limit), if we count
with appropriate theoretical models, and if we follow an optimized
scheme in the data acquisition.  Furthermore, using other instruments
like CRIRES or NIRSPEC could increase the confidence of the
results. Since our simulations exclude systematics effects specific to
the instrument, an adequate treatment to remove them would be
necessary.  Our optimal observing strategy tends to select
observations at superior conjunctions of the planet's orbit, capturing
the highest amount of light possible from the planet and at the same
time covering the highest radial velocity span for a determined time
extent.  Refinements of this technique will involve the optimization
of the distribution of time designated to the length of an observing
run \vs\ the number of nights of observation, while adding
phase-dependent functions of the planet's brightness to account for
the changing observed portion of the day/night side of the planet, and
for different amounts of irradiation in eccentric orbits, will
increase the accurateness of the fitted parameters.

\begin{acknowledgements}
  Patricio Cubillos and Patricio Rojo are supported by the FONDAP
  Center for Astrophysics 15010003, the center of excellence in
  Astrophysics and Associated Technologies (PFB06) and the FONDECYT
  project 11080271.
\end{acknowledgements}

\end{document}